\documentclass[reprint,groupedaddress]{revtex4-1}

\usepackage{graphicx}
\usepackage{dcolumn}
\usepackage{bm}%

\usepackage{hyperref}
\hypersetup{
    colorlinks=true,
    linkcolor=blue,
    filecolor=blue,      
    urlcolor=blue,
    citecolor=blue
}

\begin{document}

\title{Valley splitting depending on the size and location of a silicon quantum dot}


\author{Jonas R. F. Lima}
\email[]{jonas.de-lima@uni-konstanz.de}
\author{Guido Burkard}
\email[]{guido.burkard@uni-konstanz.de}

\affiliation{Department of Physics, University of Konstanz, 78457 Konstanz, Germany}




\begin{abstract}

The valley splitting (VS) of a silicon quantum dot plays an important role for the performance and scalability of silicon spin qubits. In this work we investigate the VS of a SiGe/Si/SiGe heterostructure as a function of the size and location of the silicon quantum dot. We use the effective mass approach to describe a realistic system, which takes into account concentration fluctuations at the Si/SiGe interfaces and also the interface roughness. We predict that the size of the quantum dot is an important parameter for the enhancement of the VS and it can also induce a transition between the disorder-dominated to deterministic-enhanced regimes. Analyzing how the VS changes when we move the quantum dot in a specific direction, we obtain that the size of the quantum dot can be used to reduce the variability of the VS, which is relevant for charge/spin shuttling. 

\end{abstract}


\maketitle

\section{Introduction}

Quantum dots (QD) in Si/SiGe heterostructures are ideal hosts for spin qubits, due to the zero-spin isotopes and the weak spin-orbit interaction \cite{RevModPhys.85.961,RevModPhys.95.025003}. Experiments have demonstrated long relaxation \cite{Morello,Yang,PhysRevApplied.11.044063} and dephasing \cite{PhysRevB.83.165301,Tyryshkin,doi:10.1126/science.1217635} times in Si/SiGe quantum dots and also high fidelities for single and two-qubit gates \cite{Veldhorst,Yoneda,doi:10.1126/science.aao5965,Watson,Huang}. More recently, a fidelity above 99\% was reached for two-qubit gates \cite{Xue,Noiri,doi:10.1126/sciadv.abn5130}, revealing the promise of this system for building high-performance and scalable qubits. However, one of the main challenges comes from the degeneracy of the conduction band minima of bulk silicon, known as valleys, which limits the performance of quantum information processing. Even though the sixfold valley degeneracy is lifted due to biaxial strain and the confinement potential in a SiGe/Si/SiGe heterostructure, the valley splittings of the two low-lying valley states, $E_{\rm VS}$, are often uncontrolled and can be very small, ranging from $\approx 10$~$\mu$eV to $100$~$\mu$eV \cite{doi:10.1063/1.3569717,doi:10.1063/1.3666232,doi:10.1063/1.4922249,PhysRevApplied.13.034068,PhysRevB.95.165429,PhysRevB.98.161404,PhysRevLett.119.176803}. With such small valley splittings, the excited state works as a leakage channel for quantum information, which poses a significant challenge for qubit operations. In this way, it is crucial to understand how the valley splitting changes as a function of the parameters of the system.

Previous work, both experimental and theoretical, has already reported, e.g., the valley splitting (VS) in a Si/SiGe quantum dots as a function of electromagnetic fields \cite{doi:10.1063/1.2387975,PhysRevResearch.2.043180}, interface roughness \cite{PhysRevB.82.205315}, well width \cite{doi:10.1063/1.1637718,PhysRevB.75.115318,PhysRevApplied.15.044033}, steps at the interface \cite{PhysRevB.100.125309,PhysRevB.94.035438,PhysRevB.104.085309}, interface width \cite{PhysRevB.80.081305}, and alloy disorder \cite{PaqueletWuetz2022,Lima_2023}. It was demonstrated, among other things, that steps can suppress the valley splitting, but only for sharp interfaces, that the electric field is an important parameter for VS engineering, while a magnetic field has a very weak effect, and that alloy disorder introduces an uncontrolled variability of the VS. The greatest values for the VS are obtained for sharp Si/SiGe interfaces ($\leq$ 2 monolayers (ML)), but the sharpest interface realized experimentally so far is 5 ML wide. This limitation together with the variability of the VS that arises due to alloy disorder led to proposals to modify  the heterostructure in order to enhance the VS, including, e.g., the introduction of a Ge layer inside of the Si well \cite{PhysRevB.104.085406} and the presence of an oscillating Ge concentration, known as wiggle well \cite{McJunkin2022,PhysRevB.106.085304}. Despite of all these works, there is no report of Si/SiGe devices engineered with reliably high valley splittings.

The most efficient way to calculate the VS is using the effective mass theory (EMT) \cite{PhysRevB.75.115318,PhysRevB.81.115324}, since the results are straightforward and analytical. Even though there are more accurate approaches, such as the NEMO-3D $sp^3d^5s^*$ model \cite{doi:10.1063/1.2591432} based on tight-biding theory, it was already demonstrated that the VS is very well described by simplified models and results obtained using the EMT are quite reliable \cite{losert2023practical}. In order to make the EMT description of the system more realistic, it was recently extended to include the alloy disorder. To this end, two models were proposed. The first one was an one-dimensional simplified model \cite{PaqueletWuetz2022} that, even though accurate, has some limitations, such as for the calculation of the influence of the lateral confinement. The second one is a three-dimensional model that incorporates microscopic features of the device in the continuum model \cite{Lima_2023}.

In this work we use the model proposed in Ref.~\cite{Lima_2023} to investigate the influence of the size and the location of a silicon quantum dot in the VS. We consider a realistic SiGe/Si/SiGe heterostructure grown in the $z$ direction, where we take into account the concentration fluctuations (alloy disorder) as well as the interface roughness. We find that the size of the quantum dot can be used to both control the distribution of the VS and enhance its average value. We find that the optimal strategy for the enhancement of the VS depends on whether the deterministic VS is above or below a defined threshold. This deterministic valley splitting is obtained when we do not take into account the alloy disorder of the device. We also determine how the VS changes when we move the quantum dot in the $xy$ plane perpendicular to the growth direction. The variation of the valley splitting depends strongly on the size of the quantum dot. This is very important, e.g., for the shuttling process \cite{cite-key,PRXQuantum.4.020305}, since the probability of an excitation from the ground state as a function of the velocity of the shuttling process depends directly on how the valley splitting changes in different locations of the device. 

The remainder of this paper is organized as follows. In Sec.~\ref{sec:model} we describe in detail the system and the model considered here, explaining how we model the concentration fluctuations and the interface roughness of the system. We also obtain the envelope function, which will be used for the calculation of the VS. In Sec.~\ref{sec:results} we obtain and discuss all the results of the work. First, we describe the influence of the size of the quantum dot in the VS, Sec.~\ref{sec:size}, and then we show the variation of the VS as we move the quantum dot on the device, Sec.~\ref{sec:location}. The paper is summarized and concluded in Sec.~\ref{sec:conclusion}. 

\section{Model}
\label{sec:model}

\begin{figure}[h]
\includegraphics[width=0.7\linewidth]{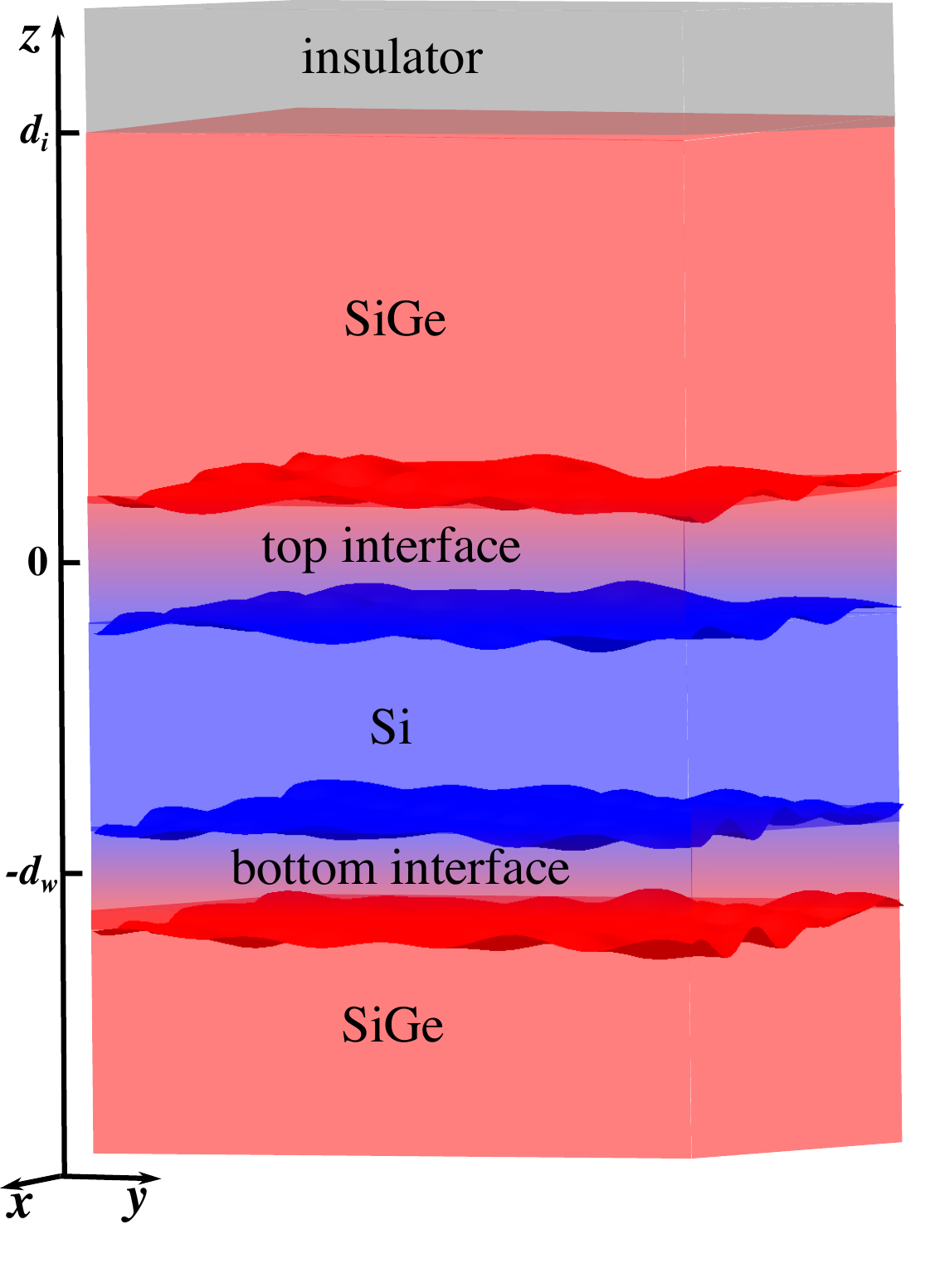}
\caption{\small Schematic diagram of the SiGe/Si/SiGe heterostructure containing the quantum dot. The device has six regions: the Si well, the top and bottom interfaces, the lower and upper SiGe barriers and a insulator layer. We take into account the interface roughness, which means that the interface width changes randomly as function of $x$ and $y$, as can be seen in the red and blue surfaces, which are the surfaces that delimit the interface regions. We consider that the potential is infinite in the insulator region.} \label{system}
\end{figure}

The system considered here is a realistic SiGe/Si/SiGe heterostructure, which is shown in the schematic diagram of Fig.~\ref{system}. The heterostructure is assumed to be grown along the $\hat{z}$ direction and a quantum well is created in the silicon region. Here, this region has a width of $d_w = 10$~nm, which is a value usually considered in the fabrication of such devices \cite{10.1063/5.0101753,Liu_2023}, and is located at $-d_w \leq z \leq 0$. We consider that the SiGe barrier regions have 30\% germanium and that the germanium concentration does not increase abruptly at the Si/SiGe interfaces, which means that the interfaces have a non-zero width. The sharpest interface width realized experimentally so far in such heterostructures was 5 monolayers (ML) wide \cite{PaqueletWuetz2022}, where 1 ML = 0.14 nm. We also take into account the interface roughness, which means that the interface width changes randomly as a function of the coordinates $x$ and $y$, as we  describe in more detail below. Additionally, electrostatic gates (not shown in Fig.~\ref{system}) are used to induce an electric field in the $\hat{z}$ direction and also to confine electrons in the silicon layer, creating a silicon quantum dot. Such quantum dots are the hosts of the spin qubits. On top of the device, we included an insulator region located at $z=d_i = 46$~nm, which is a value much greater than the small penetration of the envelope function into the upper SiGe barrier. 

Within the EMT, in order to calculate the envelope function of the system, one usually considers a confinement potential in the $\hat{z}$ direction $U(z)$, which is zero in the silicon well and, for SiGe with 30\%  germanium content, amounts to $150$~meV within the SiGe barriers, which corresponds to the energy offset between the conduction band minima of Si and SiGe. At the interfaces, this potential fluctuates due to random concentration fluctuations. It is important to mention that the VS is very sensitive to the random distribution of Si and Ge atoms within the Si/SiGe interface region and the sample-to-sample alloy disorder gives rise to a statistical distribution of VS. 

Here, we consider a model proposed recently in Ref.~\cite{Lima_2023} where the confinement potential $U(z)$ is replaced by a sum of delta functions at the location $\vec{r}_i=(x_i,y_i,z_i)$ of each Ge atom,
\begin{equation}
U(x,y,z)=\lambda \sum_i \delta(x-x_i) \delta(y-y_i) \delta(z-z_i),
\label{U}
\end{equation}
where $\lambda = 10$~meV$\cdot$nm$^3$ is a fixed parameter of the model and $i$ labels the Ge atoms. In Ref.~\cite{Lima_2023}, this potential was used only at the Si/SiGe interfaces. Here, we extend it to all regions, since the concentration fluctuations at the SiGe barriers are also relevant for the VS. One of the advantages of this model is that, e.g., the alloy disorder and the interface roughness can be easily included in the calculations. This is done by adjusting the location of the randomly distributed delta functions. Also, it is possible to calculate the VS as a function of the transversal size and location of the quantum dot directly, since this is a three-dimensional model of the system. 

The Hamiltonian that describes the envelope function of the system is given by
\begin{eqnarray}
H=\frac{p_x^2}{2m_t} + \frac{1}{2} m_t \omega_x^2 x^2 + \frac{p_y^2}{2m_t} + \frac{1}{2} m_t \omega_y^2 y^2 \nonumber \\
+ \frac{p_z^2}{2m_l} -eF_zz + U(x,y,z),
\label{H}
\end{eqnarray}
where $m_t = 0.19 \; m_e$ and $m_l = 0.98 \; m_e$ are the transverse and longitudinal effective masses, $m_e$ denotes the free electron mass, and $\omega_x = 2\hbar/m_tx_0^2$ and $\omega_y = 2\hbar/m_ty_0^2$ are the confinement frequencies along $\hat{x}$ and $\hat{y}$ directions, with $x_0$ and $y_0$ being the size (radius or semi-axis) of the quantum dot along $\hat{x}$ and $\hat{y}$. We consider in all results here a strong electric field of $F_z = 20$~MV/m, since it was demonstrated recently that the VS can be considerably enhanced by the electric field \cite{PhysRevResearch.2.043180,Lima_2023}. 

Even though we have a very powerful model that can incorporate many properties of the device and allows for a direct calculation of the VS as a function of various parameters of the system, the price to pay is that the Hamiltonian (\ref{H}) is not separable. We deal with this by replacing the potential (\ref{U}) in each realization by an average potential $\bar{U}$ plus a fluctuation $\delta U$. The average is taken over 10$^4$ realizations of the alloy disorder in all results presented here. Without interface roughness, we consider that the Ge atoms are distributed uniformly in the $x$ and $y$ directions. At this way, the average potential will be constant in these directions and we can write
\begin{eqnarray}
U(x,y,z)= \bar{U}(z)+\delta U(x,y,z).
\end{eqnarray}
Thus, if we treat $\delta U$ as a small perturbation, the unperturbed Hamiltonian becomes separable. 

When we include the interface roughness, we can no longer consider a uniform distribution in the $x$ and $y$ directions, since now the interface width and the location of the interface change as a function of these coordinates. However, for all points in the $xy$ plane, the interface width and the location of the interface fluctuate in each realization around the same value, which means that the average value is the same in all points. So, $\bar{U}$ still depends only on the $z$ direction. 

We model the average potential by considering that the Ge atoms are distributed in the $z$ direction following a probability distribution function (PDF) given by a hyperbolic tangent function. In this way, we have that
\begin{eqnarray}
\bar{U}(z)= \frac{U_0}{2}\left[\tanh ((-d_w-z)/L_b)+1\right] \nonumber \\
+\frac{U_0}{2}\left[\tanh (z/L_t)+1\right],
\label{ubar}
\end{eqnarray}
where $U_0 = 150$~meV and $L_b$ and $L_t$ control the widths of the bottom and top interfaces, respectively. The case with $L_{b(t)} = 0$ means an ideally sharp step interface. In order to avoid the presence of germanium atoms in the silicon well region, we are cutting Eq.~(\ref{ubar}) off when the argument of the hyperbolic tangent function is lower than -2.7, and hence $\bar{U}/U_0<0.01$. In this way, the probability of a germanium atom appearing at $(-d_w + 2.7)L_b < z < -2.7\cdot L_t$ is equal to zero and we can define the geometric width of the interface as $5.4\cdot L_{b(t)}$, which is the distance between the red and blue surfaces that delimit the interface regions showed in Fig.~\ref{system}. Other functions could also be used to model the average potential, such as the error function \cite{PhysRevB.80.081305,PhysRevApplied.13.044062} and the sigmoid function \cite{PaqueletWuetz2022,losert2023practical}.

\subsection{Interface roughness}

\begin{figure}[h]
\includegraphics[width=\linewidth]{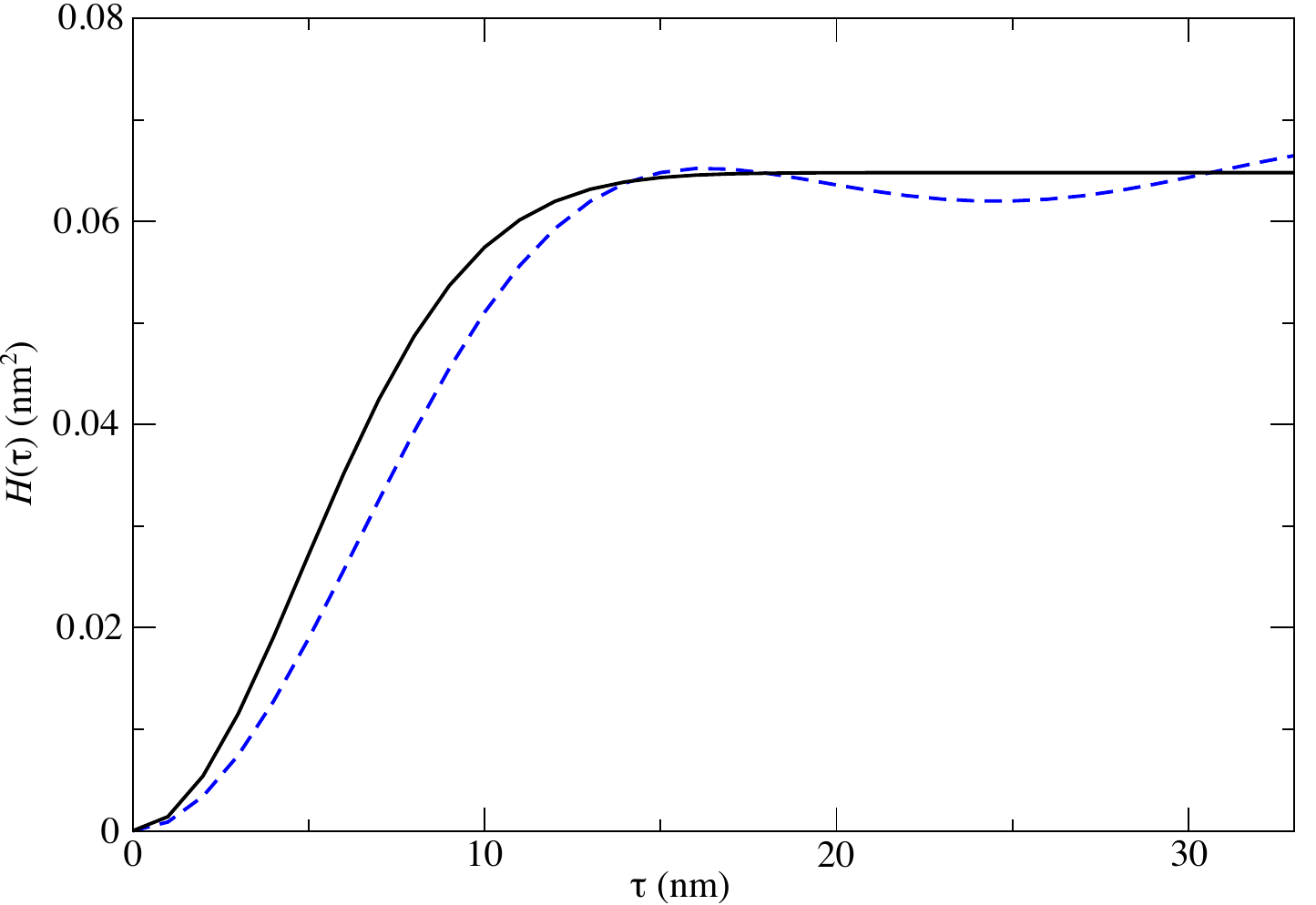}
\caption{\small Height-height correlation function $H(\tau)$ as a function of the length $\tau$ of the in-plane vector $\vec{\tau}$. Black solid line: the function $2\Delta^2[1-\exp(-(\tau/\Lambda)^2)]$, which fits the experimental data in Ref.~\cite{PhysRevApplied.13.044062}. Blue dashed line: $H(\tau)$ obtained from Eq.~(\ref{height}).} \label{rough}
\end{figure}

The interface roughness is characterized by two main length scales: (i) the root mean square (rms) fluctuations $\Delta$, which measure the vertical extent of the roughness, and (ii) the correlation length $\Lambda$, which measures its horizontal spread. Different values for $\Delta$ and $\Lambda$ were already obtained experimentally for the interface between distinct semiconductors \cite{PhysRevB.32.8171,Yoshinobu_1994,10.1063/1.1494124,PhysRevApplied.13.044062}. In particular, in Ref.~\cite{PhysRevApplied.13.044062}  the height-height correlation function $H(\tau)$ for a Ge/SiGe interface
was determined experimentally. This function is defined as
\begin{eqnarray}
H(\tau) = \left\langle |h(\vec{\rho}) - h(\vec{\rho}+\vec{\tau})|^2 \right\rangle_{\vec{\rho}},
\end{eqnarray}
where $h(\vec{\rho})$ is the height fluctuation as a function of the in-plane coordinates $\vec{\rho} = (x,y)$. So, $H(\tau)$ is the average in the in-plane direction of the squared difference of $h(\vec{\rho})$ calculated at two points separated by the in-plane vector $\vec{\tau}$. The measured $H(\tau)$ can be fitted by the function $2\Delta^2[1-\exp(-(\tau/\Lambda)^2)]$, with $\Delta=0.18$~nm and $\Lambda=6.98$~nm. We plotted this function in Fig.~\ref{rough} (black solid line).

We model the roughness by the height function
\begin{eqnarray}
h(\vec{\rho}) =\sum_{j=1}^{j_{\rm max}} a_j\cos\left(j\frac{x}{d_0}+\phi_{x_j}\right)\cos\left(j\frac{y}{d_0}+\phi_{y_j}\right),
\label{height}
\end{eqnarray}
where $a_j$, $\phi_{x_j}$ and $\phi_{y_j}$ are random parameters. If we consider $j_{\rm max} = 100$,  $1/d_0 = 0.003$~nm$^{-1}$,  $\phi_{x_j}$ and $\phi_{y_j}$ as random phases distributed uniformly in the range $[0,2\pi]$ and  $a_j$ being uniformly distributed in the range [-0.07 nm, 0.07 nm], we can use the height fluctuation (\ref{height}) to reproduce the height-height correlation function obtained in Ref.~\cite{PhysRevApplied.13.044062}, as shown in Fig.~\ref{rough} (blue dashed line). 

The interface roughness is used here to define the beginning and the end of the interface region at different points of the $xy$ plane. For instance, without interface roughness, the top interface, which is centred at $z = 0$, is located in the region $-2.7L_t < z < 2.7L_t$. With the interface roughness, this range is modified to $-2.7L_t + h_1(\vec{\rho}) < z < 2.7L_t+h_2(\vec{\rho})$. Since we are considering that $h_1(\vec{\rho})$ is different from $h_2(\vec{\rho})$, the interface width and the location of the interface will depend on $x$ and $y$ due to the roughness. One can see the interface roughness defined here in the red and blue surfaces of Fig.~\ref{system}.

We note that the roughness measured experimentally in Ref. \cite{PhysRevApplied.13.044062} was for an interface width of 24 MLs. Here, we will consider different interface widths. We expect the interface roughness to be smaller than the interface width. So, we modify the roughness proportionally. In our example, when decreasing the interface width from 24 MLs to 12 MLs, we replace $h(\vec{\rho})$ by $h(\vec{\rho})/2$.

\subsection{Envelope function}

As discussed previously, considering $\delta U$ as a perturbation, the Hamiltonian (\ref{H}) becomes separable. This implies that we can write the envelope function as a product, $\Psi_{xyz} = \psi_x\psi_y\psi_z$, where $\psi_x(x)$ and $\psi_y(y)$ are harmonic oscillator wavefunctions, which have well-known eigenenergies $E_{x,n_x}$ and $E_{y,n_y}$. The solution along the $z$ direction can be obtained as it was done in Ref.~\cite{Lima_2023}. Using the electrical confinement length
\begin{eqnarray}
z_0 = \left(\frac{\hbar^2}{2m_leF_z}\right)^{1/3},
\end{eqnarray}
and the energy scale
\begin{eqnarray}
\epsilon_0 = \frac{\hbar^2}{2m_lz_0^2},
\end{eqnarray}
the Schr\"odinger equation for the envelope function in the $z$ direction can be written as
\begin{eqnarray}
\left[ \frac{d^2}{d\tilde{z}^2} -(\tilde{U} - \tilde{z}   - \tilde{\epsilon}_{z,n_z}) \right]\psi_{z,n_z}=0,
\label{sch2}
\end{eqnarray}
where $\tilde{U} = \bar{U}/\epsilon_0$, $\tilde{z} = z/z_0$ and $\tilde{\epsilon}_{z,n_z} = \epsilon_{z,n_z}/\epsilon_0$.
The equation above has an analytical solution only when $\tilde{U}$ is constant, which is given by a linear combination of the Airy functions of the first and second kind. We solve it numerically using the transfer matrix method. We neglect in this work the corrections in the envelope function due to the perturbation $\delta U$, since these corrections would not have a relevant influence in the valley splitting. We explain it in more detail in the next section. 

\begin{figure}[h]
\includegraphics[width=\linewidth]{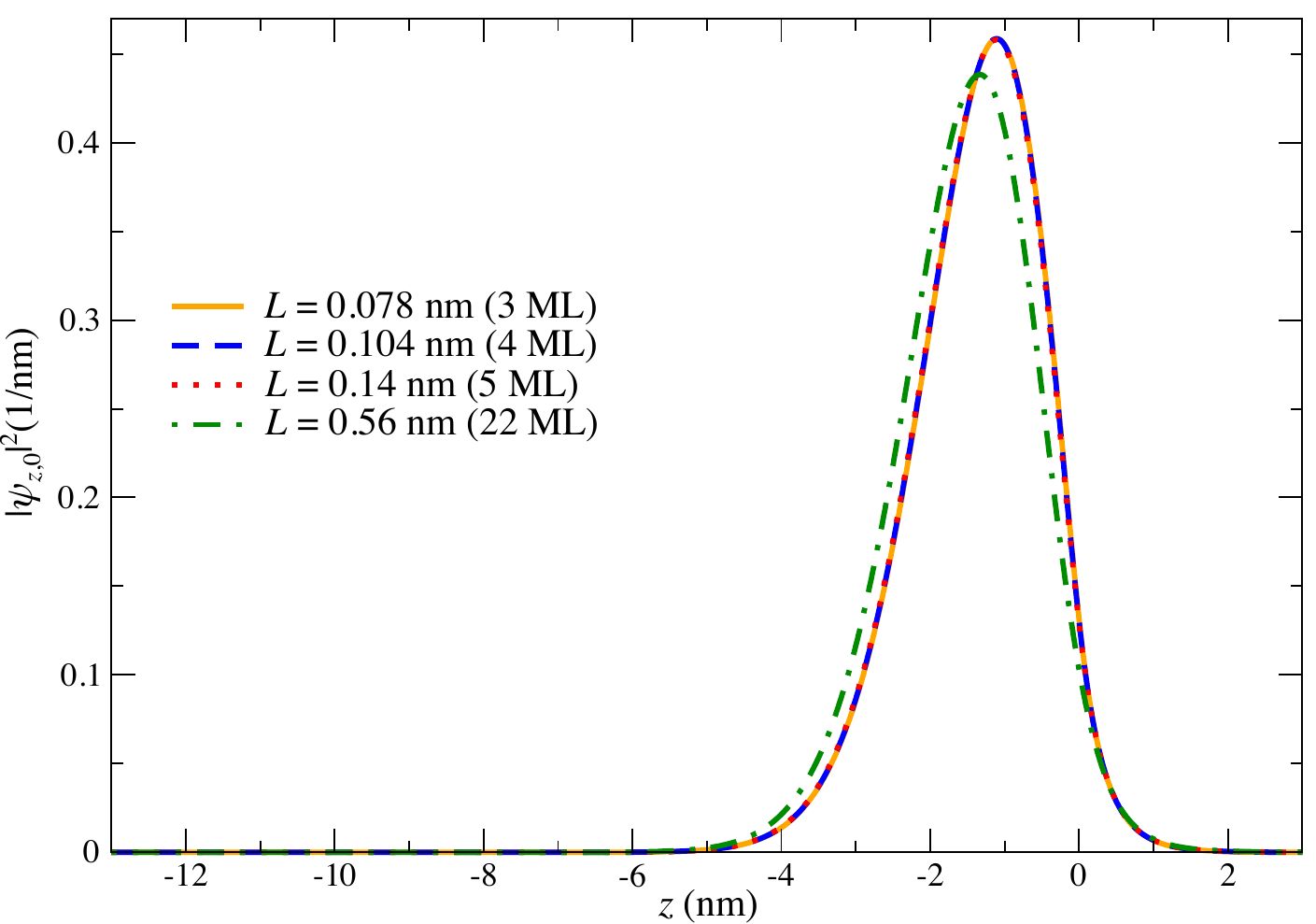}
\caption{\small Probability density along the $z$ direction $|\psi_{z,0}|^2$ for the ground state of the quantum dot for four different values of $L$ (interface width). In all curves we consider an electric field of $F_z = 20$~MV/m.} \label{envelope}
\end{figure}

In Fig.~\ref{envelope} we plot the probability density $|\psi_{z,n}|^2$ for the ground state of the quantum dot. We consider four different interface widths. One can note that when the interface width changes by a small amount (1 ML), there is no relevant change in the probability density. We can see only a small change when the interface width increases from 5 to 22 ML. This is one of the reasons that leads us to neglect in this work the corrections in the envelope function due to the perturbation $\delta U$, since the small changes in the interface width introduced by the interface roughness have almost no effect in the probability density. 

Due to the strong electric field considered here, the envelope function is pushed towards the top interface. So, for a Si well of 10 nm, the bottom interface will not contribute to the valley splitting. Also, the envelope function has only a small penetration in the upper SiGe barrier. This is relevant for the calculation of the valley splitting, since it means that we can neglect the Ge atom below the silicon well and also the Ge atoms in the upper SiGe barrier that are far from the top interface. 

 \subsection{Valley splitting}
 
 With the envelope function, we can now obtain the valley splitting. The Bloch wavefunction of the two low-lying valley states of the silicon quantum dot can be written as
\begin{equation}
|\pm z\rangle = \Psi_{xyz}e^{\pm ik_0z}u_{\pm z}(\textbf{r}),
\end{equation}
where $ \Psi_{xyz}$ is the envelope function, $u_{\pm z}(\textbf{r})$ are the periodic parts of the Bloch wavefunctions, $k_0 = 0.82(2\pi /a_0)$ is the Bloch wavenumber at the conduction band minima of silicon and $a_0 = 0.543$~nm is the length of the Si cubic unit cell. 

The intervalley coupling is given by
\begin{equation}
\Delta = \langle +z | -eF_zz+U(x,y,z)|-z\rangle.
\end{equation} 
The contribution to the intervalley coupling from the electric field is negligibly small compared to the other term. So, we can write
\begin{eqnarray}
\Delta=C_0\int e^{-2ik_0z}U(x,y,z)|\Psi_{x,y,z}|^2 d^3x,
\label{inter}
\end{eqnarray}
where $C_0 = -0.2607$ comes from the periodic parts of the Bloch wavefunctions \cite{PhysRevB.84.155320,PhysRevResearch.2.043180}. 
The total valley splitting is then given by
\begin{eqnarray}
E_{\rm VS} = 2 |\Delta|.
\end{eqnarray}

The integral in the coordinate $z$ of Eq.~(\ref{inter}) can be understood as the $2k_0$ Fourier component of the potential $U(x,y,z)$, weighted by the probability density $|\Psi_{x,y,z}|^2$. In this way, we can say that the valley splitting is a consequence of the Fourier components of the confinement potential at this wavevector. This is called the $2k_0$ theory \cite{losert2023practical}. Since the potential $U$ is determined by the Ge concentration of the device, the valley splitting is very sensitive to the distribution of the Ge atoms. It explains, e.g., the variability of the VS due to the alloy disorder and also the small VS obtained for a smooth potential and the enhancement of the VS for a sharp interface. This is also the main reason for us to neglect the corrections in the envelope function due to the perturbation $\delta U$, since the main contribution to the valley splitting comes from abrupt changes in the potential $U$ and not from small corrections in the smooth envelope function.  

Another important point from Eq.~(\ref{inter}) is that Ge atoms located in regions where the probability density is very small do not contribute to the valley splitting. Therefore, in the $z$ direction, we consider for the calculation of the valley splitting only Ge atoms in the range $-2.7L < z < 27L$. These are the Ge atoms located at the top interface and in the bottom of the upper SiGe barrier. In the $x$ and $y$ directions, we take into account only Ge atoms inside of a circle with twice of the semi-axis of the quantum dot. No other Ge atoms will have a relevant contribution to the valley splitting.

\section{Results and discussion}
\label{sec:results}

In this section we show and discuss the results obtained for the VS of the system considered here using the model described in the previous section. We divide this section in two parts. First, we consider the influence of the size of the quantum dot on the VS, and second, we analyse how the VS changes when we move the quantum dot in a shuttling process.  

\subsection{Size of the quantum dot}
\label{sec:size}

In order to study the influence of the size of the silicon quantum dot on the VS, we consider circular quantum dots where $x_0=y_0$ and consequently $\omega_x = \omega_y = \omega$, with the radius ranging from 5~nm to 25~nm. The smallest quantum dot that we consider here (5~nm radius) has a size compatible with the quantum dots obtained experimentally, e.g., in Ref.~\cite{PaqueletWuetz2022}. When increasing the size of the quantum dot, we need to avoid an exceedingly small value for the orbital splitting ($\hbar \omega$), which would create a new degree of freedom to compete with spin as the qubit two-level system, reducing the spin-qubit performance. For the biggest quantum dot considered here (25~nm radius), the orbital splitting is given by 1284~$\mu$eV, which is far from the value of the valley splitting usually obtained in such devices and also far from the Zeeman splitting that would be used for spin qubit processing. So, it reveals that we are considering here realistic radius for the quantum dots. 

\begin{figure}[t]
\includegraphics[width=\linewidth]{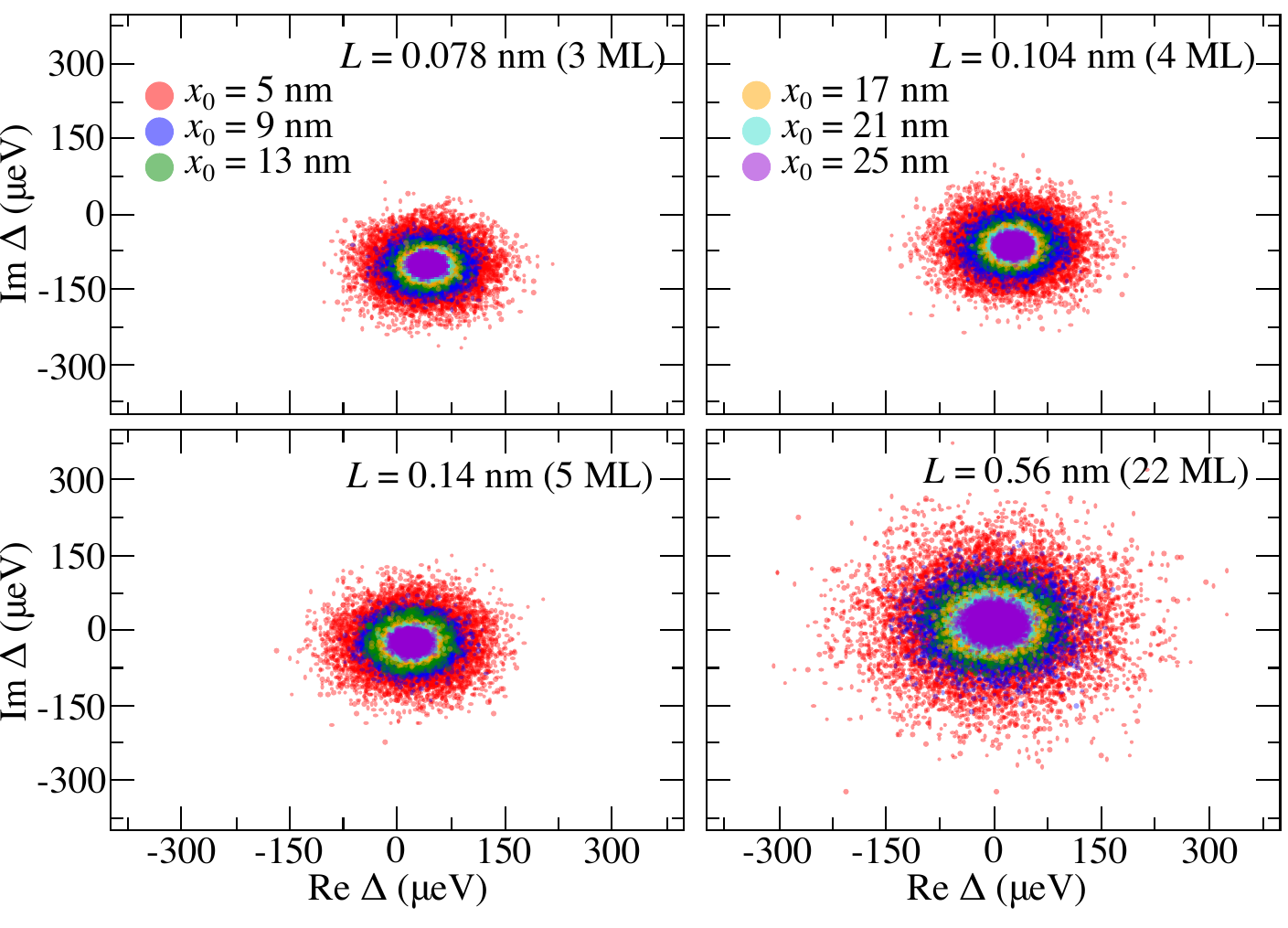}
\caption{\small The intervalley coupling $\Delta$ in the complex plane for six values of the quantum dot radius $x_0$ (color coded) and four interface widths. Each point represents one of the $10^4$ realizations considered here.} \label{RI}
\end{figure}

Initially, we consider the results without taking into account the interface roughness. In Fig.~\ref{RI} we plot the intervalley coupling $\Delta$ in the complex plane for six radii of the quantum dot and four interface widths. Each point represents one of the $10^4$ realizations considered here. We can see that the deterministic intervalley coupling $\Delta_0$, which lies in the center of the distribution, is the same for different values of the dot radius. The deterministic intervalley coupling $\Delta_0$ is obtained without considering alloy disorder, which means that it is given by Eq.~(\ref{inter}) replacing the potential $U$ by the average potential. As discussed in the previous section, the average potential depends only on the $z$ direction. For this reason, $\Delta_0$ is the same as we change the radius of the quantum dot. The main difference in the distribution of intervalley coupling when we change $x_0$ comes from the alloy disorder, which is related to the standard deviation of the valley splitting. For small quantum dots, we have greater disorder, which induces, as a consequence, a greater standard deviation. Another important point is that $|\Delta_0|$ is large for sharp interfaces (3 ML) and it reduces when we increases the interface width, which is in agreement with previous works addressed to the investigation of the valley splitting as a function of the interface width.

\begin{figure}[t]
\includegraphics[width=\linewidth]{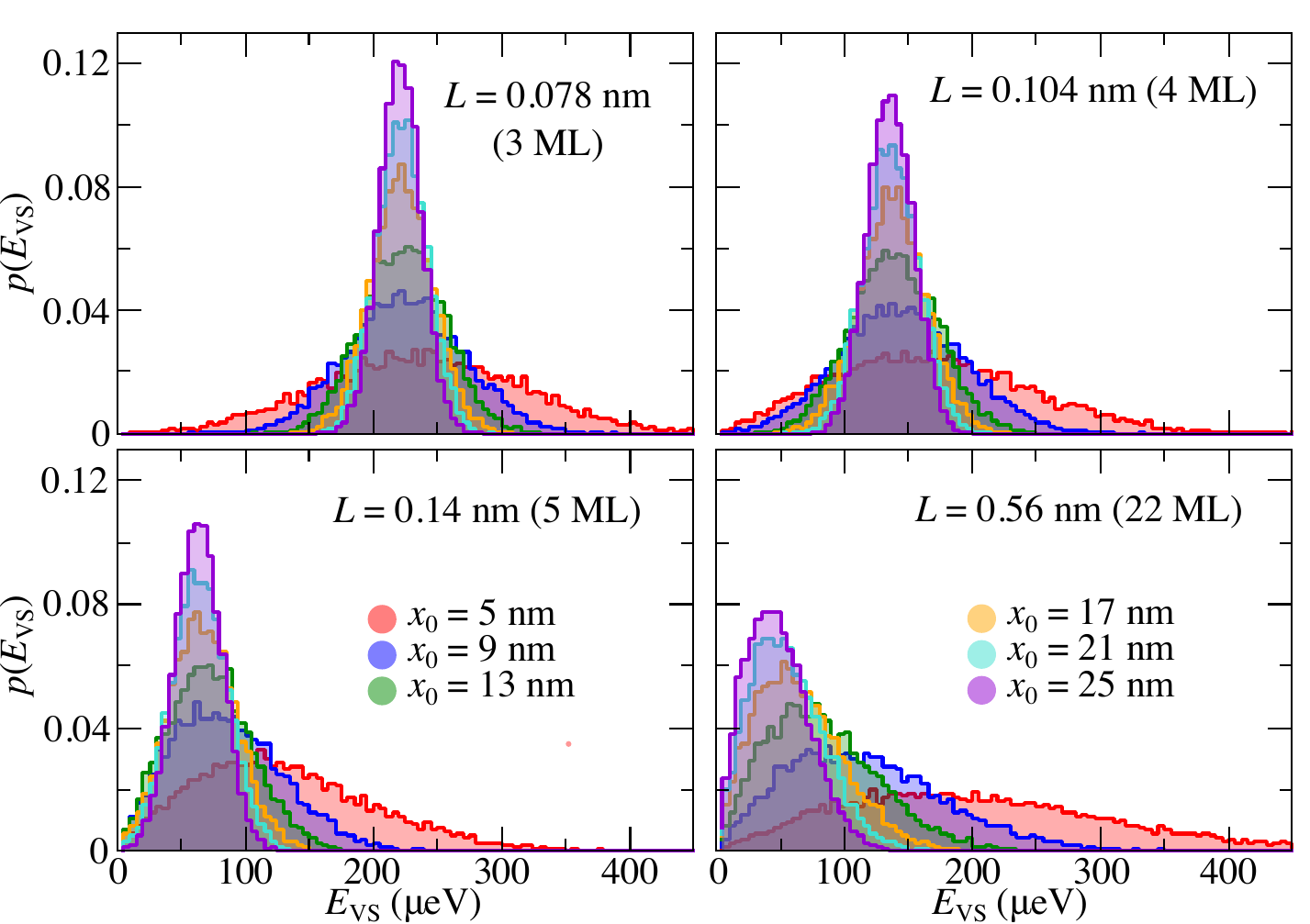}
\caption{\small Histogram of the probability density of the valley splitting. We consider here the same cases as in Fig~\ref{RI}.} \label{p}
\end{figure}

The probability density of the valley splitting $p(E_{\rm VS})$ is plotted in Fig.~\ref{p} for the same cases considered in Fig.~\ref{RI}. The circular distribution of the intervalley coupling in the complex plane leads to a Rice distribution of the VS. This distribution is characterized by the standard deviation $\sigma$ and the displacement from the origin of the circular distribution, which in our case is related to $|\Delta_0|$. As discussed in Ref.~\cite{losert2023practical}, we can define two regimes for the distribution of the valley splittings: (i) the deterministic enhanced regime, which occurs when $|\Delta_0| \gg \sigma$, and (ii) the disorder-dominated regime, which takes place when $|\Delta_0| \ll \sigma$. In our results we can see the appearance of the two regimes. For instance, for $L = 0.078$~nm (3 ML), we are in the deterministic enhanced regime, irrespective of the value of the quantum dot size considered here. This is the best situation for spin qubits applications, because in this regime, we reduce the probability of small values of the VS to appear. On the other hand, for a large interface, $L = 0.56$~nm (22 ML), we are in the disorder-dominated regime for all values of $x_0$. In this case, we always have the probability of finding devices with very small VS. In contrast to these cases, the quantum dot size can induce a transition from the deterministic enhanced to the disorder-dominated regime for the case with $L = 0.14$~nm (5 ML). This transition occurs when we increase the quantum dot radius from 9~nm to 13~nm. 

A very important point revealed in these results is the possibility of enhancement of the VS by changing the size of the quantum dot. If we define, for instance, a threshold of 100~$\mu$eV, in such a way that we would like to have as many VSs above this value as possible, the best strategy for the enhancement of VS depends only on the deterministic VS $E_{\rm VS0} = 2|\Delta_0|$, and not in which regime the distribution of the VS is. E.g., for an interface of 3 and 4 ML, the deterministic VS is above the threshold. So, in order to enhance the VS, the best strategy is to increase the size of the quantum dot. On the other hand, for an interface of 5 and 22 ML, we have that $E_{VS0}$ is smaller than the threshold. In this situation, we have to decrease the size of the quantum dot in order to enhance the VS.

\begin{figure}[t]
\includegraphics[width=\linewidth]{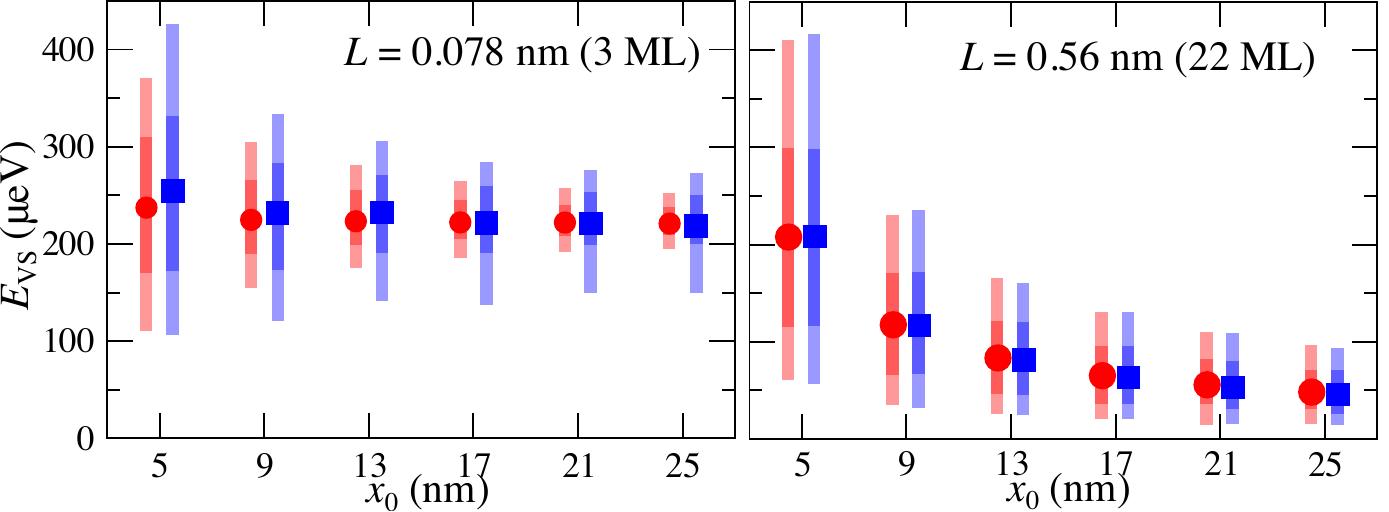}
\caption{\small The distribution of valley splittings for different values of the quantum dot radius $x_0$ with (blue markers and bars) and without (red markers and bars) interface roughness. We consider here two interface widths. The darker bars represent the 20-80 percentile rage, while the lighter bars represent the 5-95 percentile range. The circular and square markers indicate the average valley splitting.} \label{vs}
\end{figure}

The influence of the radius of the quantum dot on the valley splitting can be seen more clearly in Fig.~\ref{vs} (red markers and bars), where we consider the distribution of valley splittings for two interface widths and six quantum dot radii. The lighter bars represent the 5-95 percentile, while darker bars represent the 20-80 percentile. The red circular markers indicate the average VS. We can see that, for a sharp interface width (3 ML), the average VS does not change with the radius of the quantum dot, which is a characteristic of the deterministic enhanced regime. Looking to the distribution of VSs, the results predict more than 95\% of the devices with a VS above 100~$\mu$eV for all quantum dot radius considered here. However, for a large quantum dot (25 nm of radius), we have 95\% of the realizations with a VS above 195~$\mu$eV. On the other hand, for a wide interface (22 ML), which is in the disorder-dominated regime, the average VS decreases as we increases the radius of the quantum dot, which is a consequence of the reduction of the standard deviation. In this case, the radius of the quantum dot is an important parameter for the enhancement of the VS. For instance, we have more than 95\% of the devices with a VS smaller than 100~$\mu$eV for $x_0 = 25$~nm, while we predict more than 80\% of the realizations with a VS above 100~$\mu$eV for $x_0 = 5$~nm. 

\begin{figure}[t]
\includegraphics[width=\linewidth]{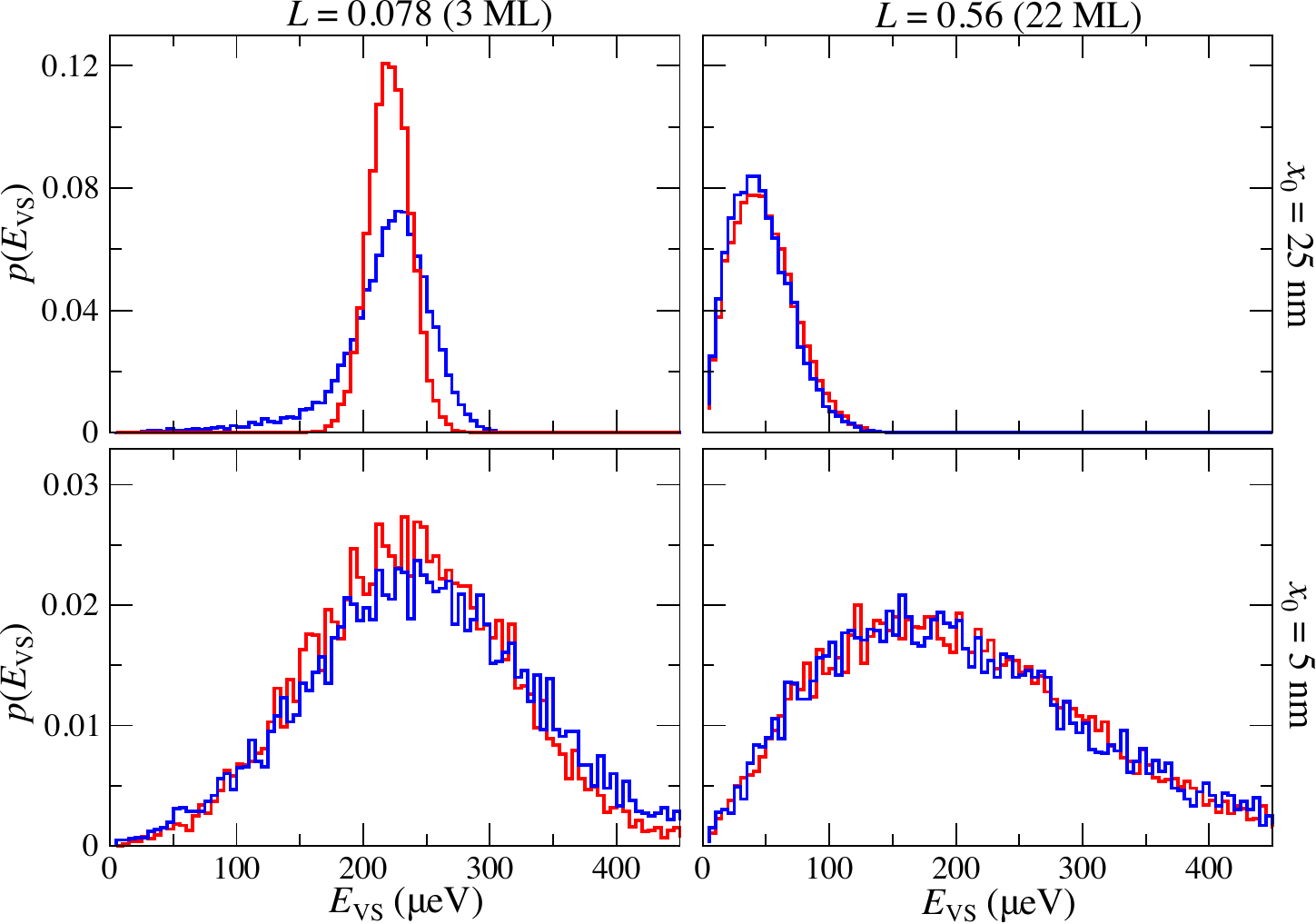}
\caption{\small Comparison of the probability density of the valley splitting between the cases with (blue curves) and without (red curves) interface roughness. We consider here two distinct interface widths and two quantum dot radii.} \label{pcomp}
\end{figure}

In Fig.~\ref{vs}, we also consider the distribution of the VS in the presence of interface roughness (blue markers and bars). For a sharp interface, the interface roughness increases the variability of the VS, which may be a consequence of the presence of more disorder in the system. However, for an interface of 22~ML, the interface roughness has no relevant influence in the distribution of the VS. In order to see this more clearly, we consider in Fig.~\ref{pcomp} the probability density of the VS with (blue curves) and without (red curves) interface roughness. The interface roughness plays an important role only when we have a sharp interface and a large quantum dot, which is the case with a weak contribution from alloy disorder. For the other cases, where we have a strong contribution from the alloy disorder, the interface roughness is irrelevant. So, we can conclude that alloy disorder overwhelms interface roughness. 

\subsection{Location of the quantum dot}
\label{sec:location}

\begin{figure}[b]
\includegraphics[width=\linewidth]{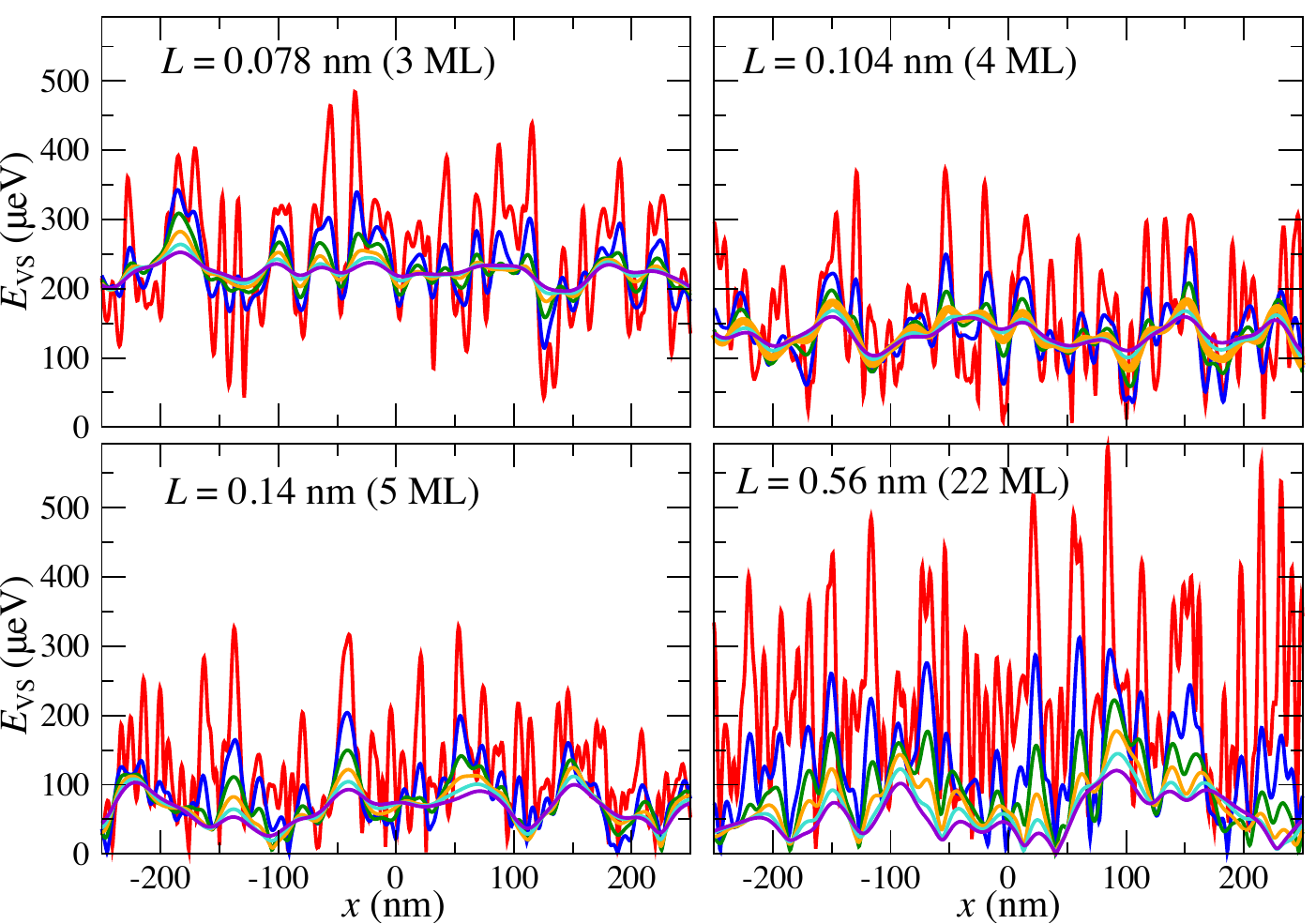}
\caption{\small The change of the valley splitting as we move the quantum dot in the $x$ direction by a distance of 500~nm. We consider four interface widths and also six values for the radius of the quantum dot (color coded). The colors used here are the same as in Figs.~\ref{RI} and \ref{p}.} \label{loc}
\end{figure}

Let us now analyze the influence of the location of the quantum dot on the VS. Here, we calculate the VS as we move the quantum dot by a distance of 500~nm in the $x$ direction, which is essentially the mechanism used in  charge and spin shuttling \cite{cite-key}. In Fig.~\ref{loc} we plot the position-dependent VS for four distinct interface widths where each color of the lines represents a different radius of the quantum dot. We used here the same colors as in Figs.~\ref{RI} and \ref{p}. 

We can see that the VS oscillates when we move the quantum dot. Looking to the amplitude of the oscillations, it is possible to note that in all cases, the VS oscillates precisely in the interval defined by the probability distribution shown in Fig.~\ref{p}.  The amplitude of the oscillations decreases as we increase the radius of the quantum dot. On the other hand, the frequency of the oscillations increases as the radius of the quantum dot increases. Therefore, in addition to a reduction in the amplitude of the oscillations, an increase of the quantum dot radius also reduces the number of visible maxima and minima of the oscillation. These results are very relevant for  charge and spin shuttling, since in the shuttling, the probability of an excitation from the ground state to the first valley excited state as a function of the shuttling velocity depends directly on how the VS changes when we move the quantum dot. Even though the calculation of this excitation probability is beyond the scope of this work, we would expect a reduction in the excitation probability in the charge and spin shuttling when we increase the radius of the quantum dot.

\begin{figure}[b]
\includegraphics[width=\linewidth]{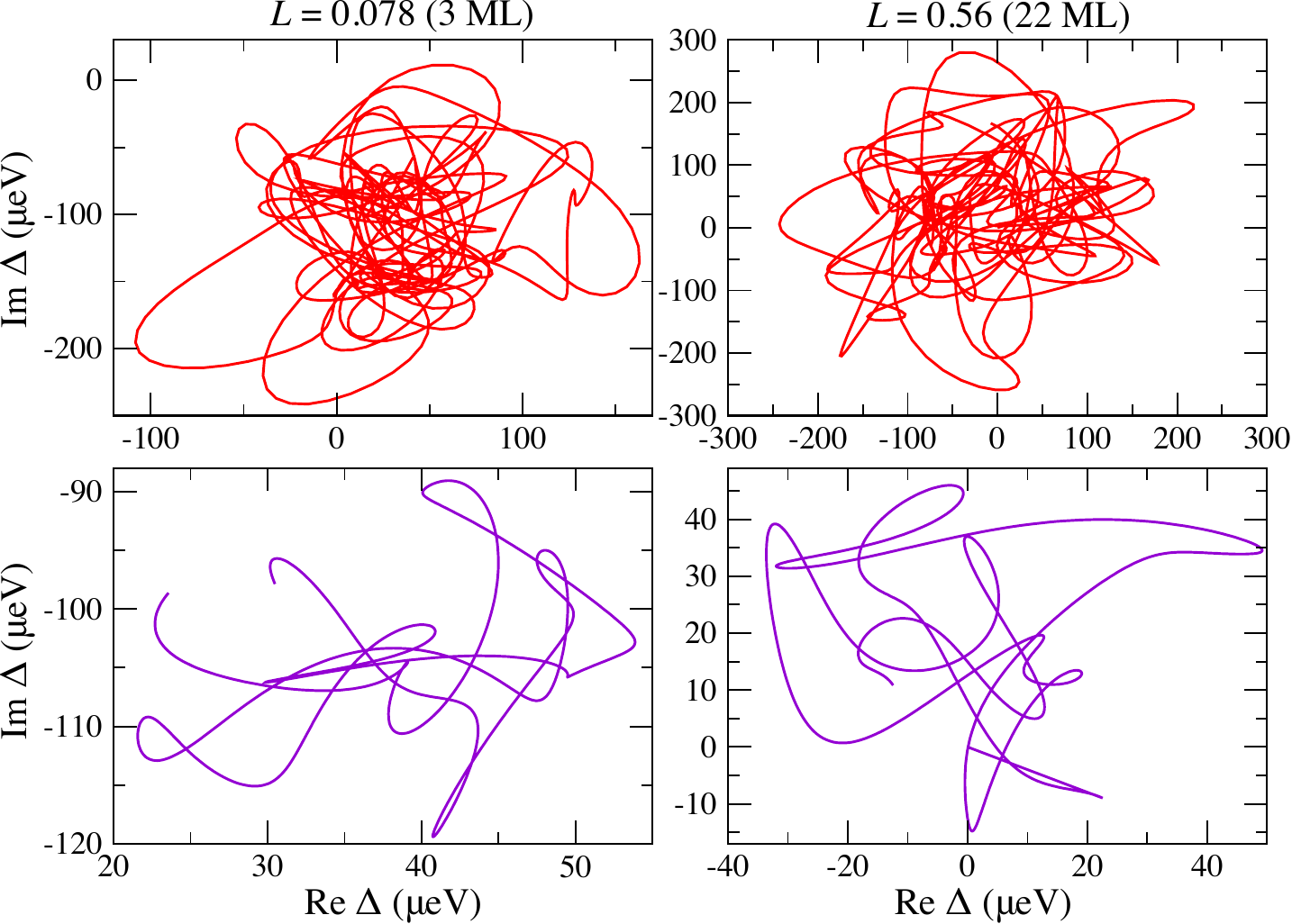}
\caption{\small The change of the real and imaginary parts of the intervalley coupling $\Delta$ when the quantum dot is moved as in Fig.~\ref{loc}. We consider here two values for the interface width and also two radii of the quantum dot. Each color represents a specific value for $x_0$ as in Figs.~\ref{RI} and \ref{p} } \label{ph}.
\end{figure}

Another relevant result is the phase $\arg(\Delta)$ of the intervalley coupling, which is also called the phase of the VS, even though only the intervalley coupling is a complex number. In Fig.~\ref{ph} we show how the real and imaginary parts of the intervalley coupling change when the quantum dot is moved as shown in Fig.~\ref{loc}. We consider two interface widths and the two colors represent two different radii for the quantum dot. We can see that, in the shuttling, the invervalley coupling moves in the complex plane in the region defined by the distribution shown in Fig.~\ref{RI}. Additionally, we note a smoother change for a bigger quantum dot radius, which is in agreement with the reduction of the number of oscillations for larger quantum dots observed in Fig.~\ref{loc}. Comparing the two distinct interface widths, we can see that for a wider interface (22 ML in this case), we have a greater change in the VS phase, since in this case the intervalley coupling moves in the four quadrants of the complex plane, which is not the case for a sharp interface. 
 
\section{Conclusion}
\label{sec:conclusion}

The valley splitting plays an important role for the performance and scalability of silicon spin qubits. Therefore, it is crucial to understand the behaviour of the VS as a function of the parameters of the system. In this work, we calculated the VS as a function of the radius and location of a quantum dot in a Si/SiGe heterostructure. Within the effective mass theory, we modeled a realistic system taking into account the alloy disorder and the interface roughness. We found that the alloy disorder overwhelms the interface roughness. So, the interface roughness will have a relevant contribution to the VS only when there is a weak influence of the alloy disorder, which is the case when we have a sharp interface and a large radius of the quantum dot. Analyzing how the VS changes as a function of the size of the quantum dot, we found that the distribution of VS is very sensitive to the radius of the quantum dot $x_0$, specially when we have a wide interface width. With an interface width of, e.g., 22 ML, we predict more than 95\% of the devices with a VS smaller than 100~$\mu$eV for $x_0 = 25$~nm, while we obtained more than 80\% of the realizations with a VS above 100~$\mu$eV for $x_0 = 5$~nm. Thus, changing the dot size can be used as a strategy for the enhancement of the VS. We also investigated how the VS changes when we move the quantum dot, which is used, for instance, for the charge and spin shuttling. We found that the VS oscillates during the shuttling. The amplitude of these oscillations can be suppressed by increasing the size of the quantum dot, which also reduces the number of oscillations. These results could be used, e.g., to predict the probability of an excitation from the ground state to the first excited valley state as a function of the shuttling velocity, which is very relevant for the performance of the shuttling process. 

\begin{acknowledgments}

This work has been funded by the Federal Ministry of Education and Research (Germany), funding code Grant No.~13N15657 (QUASAR).

\end{acknowledgments}

%

\end{document}